\documentclass[showpacs,reprint,nofootinbib,longbibliography,superscriptaddress]{revtex4-1}

\usepackage{graphicx}

\usepackage{amsmath}
\usepackage{amsfonts}
\usepackage{amssymb}

\newcommand{\tr}{\mathrm{tr}\,}

\newcommand{\Regensburg}{Institute for Theoretical Physics, Universit\"at Regensburg, D-93040 Regensburg, Germany.}

\begin{document}

\title{Diagrammatic representation of scalar QCD\\ and sign problem at nonzero chemical potential}

\author{F.~Bruckmann}
\affiliation{\Regensburg}

\author{J.~Wellnhofer}
\affiliation{\Regensburg}

\begin{abstract}
 We consider QCD at strong coupling with scalar quarks coupled to a chemical potential. Performing the link integrals we present a diagrammatic representation of the path integral weight. It is based on mesonic and baryonic building blocks, in close analogy to fermionic QCD. Likewise, the baryon loops are subject to a manifest conservation of the baryon number. The sign problem is expected to disappear in this representation and we do confirm this for three flavors, where a scalar baryon can be built and thus a dependence on the chemical potential occurs. For higher flavor number we analyse examples for a potential sign problem in the baryon sector and conjecture that all weights are positive upon exploring the current conservation of each flavor.   
\end{abstract}


\maketitle

\section{Introduction}

Diagrammatic Monte Carlo methods are one of the most successful approaches to the sign problem of quantum field theories at nonzero chemical potential, e.g., see \cite{Chandrasekharan:2008gp,
Gattringer:2014nxa,Bruckmann:2015sua}. In lattice QCD at strong coupling this idea is rather old \cite{Rossi:1984cv,Karsch:1988zx}: expanding the quark weight and integrating out the gauge links\footnote{Integrating out the quarks instead gives the determinant of the non-hermitian Dirac operator and, therefore, a complex weight.} leads to certain building blocks along the lattice bonds. They are of mesonic and baryonic nature and (together with fermion saturation at every site) facilitate an intuitive diagrammatic representation of the QCD path integral. 

Nonetheless, positivity of the so-obtained weight is not guaranteed and terms of opposite sign indeed appear, even at zero chemical potential. This has hampered further use of this approach in simulations of realistic QCD even though worm algorithms have proven capable of simulating such constrained systems (for recent attempts see \cite{Fromm:2010lga,deForcrand:2014tha}).

One may view the problem of this approach as a fermionic sign problem. For staggered fermions there are clearly four sources of negative signs: (i) the relative sign between the hopping terms in the Dirac operator which is of first order in derivatives, (ii) antiperiodic boundary conditions in Euclidean time, (iii) the Grassmann nature of the quark fields in the path integral, (iv) the staggered signs. All of them are absent if quarks were Lorentz scalars, and one may expect scalar QCD (sQCD) to be free of the sign problem. Note, however, that the gauge links are complex and SU(3) group integrals are not necessarily positive \cite{Creutz:1984mg}. To analyse the sign problem in sQCD is the main motivation of this paper. 

Gauge theories with scalar matter might be relevant beyond the Standard Model; here we compare sQCD to QCD at nonzero chemical potential. One of the main differences is the flavor-antisymmetric nature of the baryons of sQCD; as a consequence at least three scalar quark flavors are necessary to generate a dependence on the chemical potential (see Sec.~\ref{sec_sign_flavor} below). For the first interesting case of three flavors we are able to prove 
that the path integral weight is positive, i.e., this representation solves the sign problem at nonzero chemical potential.

Scalar quarks are not subject to the Pauli exclusion principle and thus the building blocks of sQCD diagrams come with less constrained occupation numbers, e.g., baryon worldlines may intersect. This shall be of advantage for numerical simulations as well as for treating higher flavor numbers, which we conjecture to be free of the sign problem as well.

\section{Original action and derivation of building blocks}

We treat sQCD with $N_f$ masssive flavors coupled to the same chemical potential $\mu$ in the strong coupling limit, i.e., without gauge (plaquette) action. The corresponding Euclidean lattice action  is the (negative) discretized SU(3) gauge-covariant Laplacian,
\begin{align}
 S
 =
 \sum_{x,f}\Big(
 &-
 \sum_\nu\big(
 e^{\mu\delta_{\nu,0}}\phi^f(x)^{\dagger} U_\nu(x) \phi^f(x+\hat{\nu})\notag\\
 &\qquad +e^{-\mu\delta_{\nu,0}}\phi^f(x+\hat{\nu})^{\dagger} U_\nu^\dagger(x) \phi^f(x)
 \big)\notag\\
 &+(2d+m^2) |\phi^{f}(x)|^2
 \Big)\,,
\end{align}
where $x$ denotes the lattice sites, $\nu=1,\ldots.,d$ is the direction index ($\nu=0$ is the temporal direction in which $\mu$ acts), $\hat{\nu}$ its unit vector and $f=1,\ldots,N_f$ is the flavor index; the lattice spacing has been set to unity. 

At real $\mu$ the action is not real, since the second line is not the complex conjugate of the first line, which is the case at $\mu=0$ (or imaginary $\mu$). In \cite{Bruckmann:2016fuj} we have presented numerical evidence that reweighting in the conventional approach of integrating out the quarks to an inverse determinant suffers from a sign problem in the sense of an oscillating phase. In particular, this gives rise to a reweighting factor, $r\sim e^{-V\Delta f(\mu)}$, that decays with the volume. 

The diagrammatic representation of this system emerges after integrating out all gauge links. To do so for a particular link, we collect the two terms in which $U_\nu(x)$ or $U_\nu^\dagger(x)$ appear and write the matter bilinears to which they couple as matrices
\begin{align}
 \begin{split}
 \sum_f
 \phi^f(x+\hat{\nu})\phi^f(x)^\dagger&
 =:J_\nu(x)\,,\\
 \sum_f
 \phi^f(x)\phi^f(x+\hat{\nu})^\dagger&
 =J^\dagger_\nu(x)\,,
 \end{split} 
\label{eq_def_J} 
\end{align}
involving an outer product in color space.
Now the gauge dependent terms in the action read
\begin{align}
 -S[U_\nu(x)]=
 e^{\mu\delta_{\nu,0}}\,\tr J_\nu(x) U_\nu(x)
 +e^{-\mu\delta_{\nu,0}}\,\tr J^\dagger_\nu(x) U_\nu^\dagger(x).
\end{align}
Note that under local gauge transformations, under which the link becomes $\Omega(x)U_\nu(x)\Omega^\dagger(x+\hat{\nu})$, the matter matrix transforms complementary, it becomes $\Omega(x+\hat{\nu})J_\nu(x)\Omega^\dagger(x)$, such that the traces in the action are gauge invariant\footnote{For fermionic quarks $J_\nu(x)$ and $J_\nu^\dagger(x)$ are commuting Grassmann bilinears, such that the presented analysis can be used for them as well, with the main difference being that powers of $J_\nu(x)$ and $J_\nu^\dagger(x)$ higher than $3N_f$ vanish due to the Grassmann nature.}.

The integration over SU(3) group elements (with Haar measure) can be turned into a five-fold sum \cite{Eriksson:1980rq}\footnote{We have further expanded the two terms $\det m+\det m^\dagger$ in \cite{Eriksson:1980rq} separately, with powers $n$ and $\bar{n}$. Note a typo in the definition of $Y$ in that reference.}
\begin{align}
 &\int\limits_{SU(3)}\!\!\!\!dU\,
 e^{\,\alpha\, \tr JU+\alpha^{-1}\, \tr J^\dagger U^\dagger}\label{eq:int_group_one}\\
 &=2\!\!\!\sum_{j,k,l,n,\bar n = 0}^{\infty}\,
 \frac{\alpha^{3(n-\bar{n})}}{g_{(1)}!g_{(2)}!}\,
 \frac{X^jY^kZ^l\Delta^{n}(\Delta^*)^{\bar n}}
 {j!k!l!n!\bar{n}!}\notag
\end{align}
over gauge invariants
\begin{align}\begin{split}
 X&= \tr(J^\dagger J)\,,\:
 Y=\frac{1}{2} \big( X^2 - \tr[(J^\dagger J)^2] \big)\,,\:
 Z=\det(J^\dagger J)\,,\\
 \Delta &= \det J\,,\:\:
 \Delta^*= \det J^\dagger=(\det J)^* \,,
\end{split}\end{align}
where
\begin{align}\begin{split}
 g_{(1)}&= k+2l+n+\bar n + 1\,,\\
 g_{(2)}&= j+2k+3l+n+\bar n +2
\end{split}\end{align}
are positive integers. When using these formulas in sQCD, one has to reinsert indices ${}_\nu$ and arguments $(x)$ on both the fields $J^{(\dagger)}$ and consequently $\{X,\ldots,\Delta^*\}$ and on the integers (`dual variables') $\{j,\ldots,\bar{n}\}$. 
The fugacity factor $\alpha=e^{\,\mu\delta_{\nu,0}}$ is present when the bond is in the 0-direction, its exponent is the difference of powers of $\Delta$ and $\Delta^*$.  
Thus the partition function $\mathcal{Z}=\int \mathcal{D}\phi \mathcal{D}U\, e^{-S}$ can 
in the diagrammatic formulation be written as
\begin{align}
 \label{eq:partition_function_dual}
 \mathcal{Z}
 =&
 \sum_{\{j,k,l,n,\bar n\}}
 \int \mathcal{D}\phi\, e^{-(m^2+2d)\sum_{x,f} |\phi^f(x)|^2}\\
 &\prod_{x,\nu} 
 \left(
 2\frac{\alpha^{3(n-\bar{n})}}{g_{(1)}!g_{(2)}!}\,
 \frac{X^jY^kZ^l\Delta^{n}(\Delta^*)^{\bar n}}
 {j!k!l!n!\bar{n}!}
 \right)_{\nu}(x),\notag
\end{align}
where the sum goes over all admissable configurations of the variables $\{j,\ldots,\bar n\}$,
to be specified more concretely in Sec.~\protect\ref{sec_sign_flavor}.

When interpreting $J$ as the hopping of (all flavors of) quarks along a bond in the direction $\nu$, then $J^\dagger$ is the hopping of (all) antiquarks on the same bond. Thus $X$ represents the hopping of `mesons' (with any pair of quark/antiquark flavors). Consistently, $X$ does not contribute a $\mu$-factor (the exponent of $\alpha$ does not contain $j$). $Y$ and $Z$ are of similar nature with two/three quarks and antiquarks hopping on a bond and no $\mu$-contribution. Therefore, we will call $X$, $Y$ and $Z$ \textit{mesonic building blocks}.

\begin{figure}[b!]
 \center
 \includegraphics{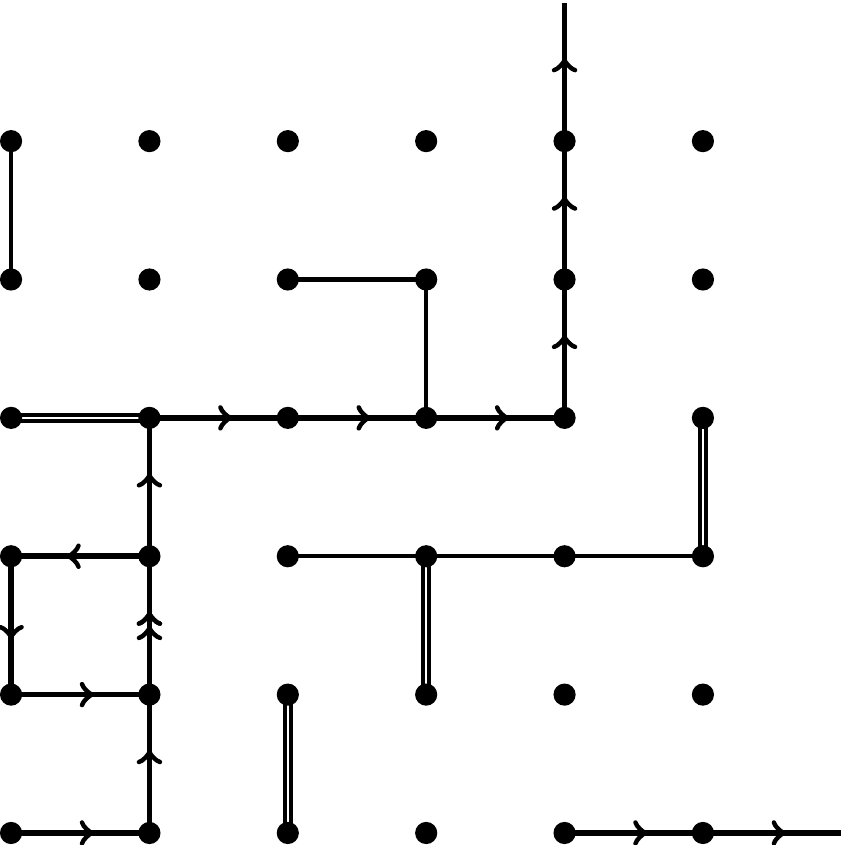}
 \caption{Example of a diagram on a $6\times 6$ lattice (with periodic boundary conditions). Unoriented single and double lines denote bonds with unit occupation of the mesonic building blocks $X$ and $Y$, i.e., $j_\nu(x)=1$ and $k_\nu(x)=1$. Oriented bonds (due to the current conservation discussed in Sec.~\protect\ref{sec_sign_flavor}) stand for baryon building blocks $\Delta$ and $\Delta^*$: arrows upwards and to the right denote $n_\nu(x)=1$ (on one bond $n_\nu(x)=2$) and arrows downwards and to the left denote $\bar{n}_\nu(x)=1$. The baryon loop winds once (actually in both directions) and thus obtains a fugacity factor $e^{3\mu/T}$ in the weight.}
\label{fig_big_example}
\end{figure}

In $\Delta$ and $\Delta^{*}$ three quarks or  three antiquarks are hopping on a bond, respectively, which is why we will call them \textit{baryonic building blocks}. As expected, they contribute positive and negative multiples of  the baryon chemical potential, $3\mu$, in the exponent. 

As is typical for bosonic systems, occupation numbers are unbounded from above and do not exclude each other. A configuration in this new representation can easily be determined by a list of all integers $\{j,\ldots,\bar{n}\}_\nu(x)$ on all bonds plus the values of the matter fields on all sites (since we have not integrated out the latter). In a visualization of these numbers one uses building blocks very similar to those from the fermionic case, see Fig.~\ref{fig_big_example}: unoriented one-, two- and three-bonds for the occupation numbers of the mesons and directed bonds for the (anti)baryons (arrows are connected to current conservation to be derived in the next section). Since multiple occupation numbers per bond are harder to visualize, the example diagram shown in that figure mostly contains single occupation numbers. Note that sites without any occupied bond are admissible, too.

As numbers, the mesons $X$, $Y$ and $Z$ are positive functions of the positive matrix $JJ^\dagger$. The factorials in Eq.~\eqref{eq:int_group_one} and the remaining Gaussian factors in Eq.~\eqref{eq:partition_function_dual} for $\phi$ are positive as well. Thus, a potential sign problem can only come from the (anti)baryons $\Delta^{(*)}$, as in fermionic QCD. One of the main features of the diagrammatic representation is that the chemical potential appearing through the fugacity $\alpha$ does not introduce signs\footnote{Interestingly, imaginary $\mu$'s, which do not induce a sign problem in the original formulation, do so in the diagrammatic representation.}, i.e., if the system has no sign problem at vanishing $\mu$ it does not develop a sign problem at nonzero $\mu$. This is in very close analogy to the defining energy representation of the grand canonical partition function.

\section{Positivity of the weight depending on the number of flavors} 
\label{sec_sign_flavor}

The objects potentially inducing a sign problem in the diagrammatic representation of sQCD are $\Delta^{(*)}=\det J^{(*)}$ and powers thereof. Importantly, the complex matrices $J$ are built out of outer products, see Eq.~\eqref{eq_def_J}, which will be analysed now. 

For  $N_f=1$ obviously any row (or column) of $J$ is linearly dependent on any other row. Consequently, the determinant of $J$ vanishes and so do all $\Delta^{(*)}$'s, such that no dependence on $\mu$ can emerge  (only $n\equiv \bar{n}\equiv 0$ contributes). 
Similarly, for $N_f=2$ at most two rows of $J$ can be linearly independent and its determinant vanishes again. We conclude that at strong coupling \textbf{sQCD develops a dependence on $\mu$ only for $\mathbf{N_f\geq 3}$}. The latter is the matrix size of $J$ and thus generalizes to the number of colors in gauge theories with higher gauge group.

For arbitrary $N_f$ the following formula is useful
\begin{align}
 &\det_3 \Big(\sum_{f=1}^{N_f} \phi^f(x+\hat{\nu}) \phi^f(x)^\dagger\Big)\notag\\
 &=\frac{1}{3!}\sum_{f_1,f_2,f_3=1}^{N_f}
 d_{f_1f_2f_3}(x+\hat{\nu})\,d^*_{f_1f_2f_3}(x)\notag\\
 &=\frac{1}{3!}\:\sum_{\sigma}
 d_{\sigma_1\sigma_2\sigma_3}(x+\hat{\nu})\,d^*_{\sigma_1\sigma_2\sigma_3}(x)\,
 \label{eq_the_formula}
\end{align}
with determinants
\begin{align}
 d_{f_1f_2f_3}(x)
 &:=\det_3\big(\phi^{f_1}(x)|\phi^{f_2}(x)|\phi^{f_3}(x)\big)\,,
\end{align}
where $|$ is used to separate three columns in a three-by-three matrix. $\sigma$ denotes choices of three flavors
\begin{align}
 \sigma
 &:\{1,2,3\}\to\{1,\ldots, N_f\}\,.
\end{align}
This formula\footnote{Note that this formula is of the type `determinant of a sum is a sum of determinants' (!) which holds for the outer product structure in which we are interested here.} (and its obvious generalization to $N_c\neq 3$) can easily be shown through writing the determinant with Levi-Civita symbols. It makes manifest the antisymmetry of the flavor indices $\{f_1,f_2,f_3\}$ (or $\{\sigma_1,\sigma_2,\sigma_3\}$) in the determinant. In the language of sQCD this means antisymmetry of quark flavors which hop together in $\det J^{(\dagger)}=\Delta^{(*)}$ representing an (anti)baryon. It also confirms the discussion above, that this determinant vanishes for less than 3 flavors. For $N_f=3$ Eq.~\eqref{eq_the_formula} has just one summand
\begin{align}
 \label{eq:baryonic_weight_nf3}
 \Delta_\nu(x)
 =\det J_\nu(x)
 &=\det\big(\phi^1(x+\hat{\nu})\big|\phi^2(x+\hat{\nu})\big|\phi^3(x+\hat{\nu})\big)\notag\\
 &\times \det\big(\phi^1(x)\big|\phi^2(x)\big|\phi^3(x)\big)^*\,,
\end{align}
where all the three quark flavors enter together just once.

So far we have not performed the matter field integrations. The $\phi$-dependent path integral weight consists of a Gaussian term, which suppresses large absolute values of $\phi$, multiplied by the product in the second line of Eq.~\eqref{eq:partition_function_dual}. The latter part is a complicated function of $\phi$. One could leave the $\phi$-integrations to numerics, provided the admissible configurations have non-negative weights.

However, there is one important feature of the $\phi$-integration which can easily be utilized; schematically
\begin{align}
 \int_{\mathbb{C}} \!d\phi\,e^{-\#|\phi|^2} (\phi)^A (\phi^*)^B\sim \delta_{AB}\,,
 \label{eq_phi_int}
\end{align}
which comes from the integration over the phase\footnote{Phase integrations typically cause U(1) current conservation in diagrammatic approaches to bosonic systems, for fermionic systems this role is played by the saturation of Grassmann integrals by equal numbers of $\psi$'s and $\bar{\psi}$'s.}  of $\phi$.
For sQCD this formula means that only those terms contribute, for which the power of $\phi^f_a(x)$ 
matches the power of its complex conjugate $\phi^f_a(x)^*$. This has to hold for every flavor $f$, color $a$ and site $x$ separately, $A^f_a(x)\stackrel{!}{=}B^f_a(x)$. 

To derive an immediate consequence for the diagrams, consider the \textbf{coarser constraints} that occur after summing these constraints over all indices but the site, $\sum_{f,a}A^f_a(x)\stackrel{!}{=}\sum_{f,a}B^f_a(x)$ for all $x$. Mesonic contributions are functions of $J^\dagger J\sim \phi_a^f(x)\phi_b^g(x)^*$ and thus contribute equal integers to both sums. The baryons $\Delta^{(*)}$, on the other hand, are of third order in $\phi^f_a(x)^*$ and $\phi^f_a(x+\hat{\nu})$, where the two factors live on neighboring sites. A single (anti)baryon thus vanishes under the $\phi$-integration and needs to be accompanied by other (anti)baryons connecting to $x$ and $x+\hat{\nu}$. One easily obtains the constraints 
\begin{align}
 \sum_\nu \big[m_\nu(x)-m_\nu(x-\hat{\nu})\big]=0\,,
\end{align}
where 
\begin{align}
 m_\nu(x)=n_\nu(x)-\bar{n}_\nu(x)\,.
\end{align}
This is nothing but the discrete version of a \textbf{manifest current conservation} $\sum_\nu\partial_\nu m_\nu(x)=0$, namely for the net baryon current $(n-\bar{n})_\nu$ (as in fermionic QCD) from all flavors. Diagrammatically, baryon building blocks must come in \textbf{closed loops}. 
In Fig.~\ref{fig_big_example}, for instance, the baryon content can be viewed as one long and winding loop plus one plaquette loop touching each other at one bond.

According to Eq.~\eqref{eq:int_group_one}, $3\mu$ couples to all  $n-\bar{n}$'s in the $0$-direction, i.e., to $\sum_x m_0(x)$, which is just the conserved charge\footnote{Since $m_\nu$ is conserved one can replace $\sum_x m_0(x)$ by $N_0\sum_{\vec{x}} m_0(x_0,\vec{x})$ (for any $x_0$) which turns $3a\mu$ with $a$ the lattice spacing into the expected factor $3\mu a N_0=3\mu/T$ with $T$ the temperature.} of this current. Equivalently, $3\mu/T$ couples to the net winding number of baryon loops in the $0$-direction. 

Coming back to the sign problem at $N_f=3$, any baryonic factor $\det(\phi^1(x)\big|\phi^2(x)\big|\phi^3(x))$ has to be accompanied by just its complex conjugate from a neighboring baryonic hopping, cf.\ Eq.\ \eqref{eq:baryonic_weight_nf3}, such that one obtains a product of positive terms\footnote{One might argue that a negative sign occurs upon permuting the flavors under one of the determinants, but according to Eq.~\eqref{eq_the_formula} this would also permute the flavors at a neighboring site and thus keep a positive sign of the total weight.} $|\det(\phi^1(x)\big|\phi^2(x)\big|\phi^3(x))|^2$ (and powers thereof) for all sites $x$ on baryon loops. This \textbf{solves the sign problem at $\mathbf{N_f=3}$}.

In the diagrammatic representation this system can therefore be simulated, presumably with a hybrid approach for the updates: for unconstrained variables such as $j_\nu(x),\,k_\nu(x),\,l_\nu(x), n_\nu(x)+\bar{n}_\nu(x)$ and $\phi^f(x)$ local updates can be used, while
for the constrained variables $n_\nu(x)-\bar{n}_\nu(x)$ worm algorithms are promising.

We close by discussing the technicalities faced at more than three flavors, say at $N_f=4$. The baryonic matching described above for $N_f=3$ does not work here:  according to Eq.~\eqref{eq_the_formula} $d_{123}$ from one baryon factor multiplies not only $d_{123}^*$ from another baryon factor, but the sum $\# d_{123}+\# d_{124}+\# d_{134}+\# d_{124}$ multiplies the sum $\# d_{123}^*+\# d_{124}^*+\# d_{134}^*+\# d_{124}^*$ with the factors $\#$ all different (determined by the fields at two neighboring sites). Obviuosly there are mixed terms, for which no positivity argument applies. Indeed, the configuration which just contains one closed baryon loop has a complex weight generically.

It seems necessary to explore \textbf{finer constraints} than above. For instance, the summand $d_{123}(x)d_{124}^*(x)$ has a `mismatch' in that it contains $\phi^3(x)$ and $\phi^4(x)^*$ once, but not their complex conjugates. This summand thus vanishes under the $\phi^3(x)$- or $\phi^4(x)$-integration, cf.\ Eq.~\eqref{eq_phi_int}. In the absence of other occupied bonds connecting to it, each baryon loop thus contains only terms $|d_{f_1,f_2,f_3}|^2$ and, therefore, is positive. 

At first sight, mesons may change this positivity argument. A mesonic building block, say $X$, connecting to the baryon loop at site $x$ does contain the `missing' factor $\phi^3(x)^*\phi^4(x)$ (with arbitrary gauge indices) to make the summand $d_{123}d_{124}^*$ survive the $\phi$-integration. However, this summand in $X$ also carries $\phi^3(y)\phi^4(y)^*$ at a neighboring site $y$. The $\phi$-integration at $y$ thus gives zero. As a consequence, the baryon loop reduces to terms $|d_{f_1,f_2,f_3}|^2$ multiplying the positive mesonic weight.

\begin{figure}
 \center
 \includegraphics{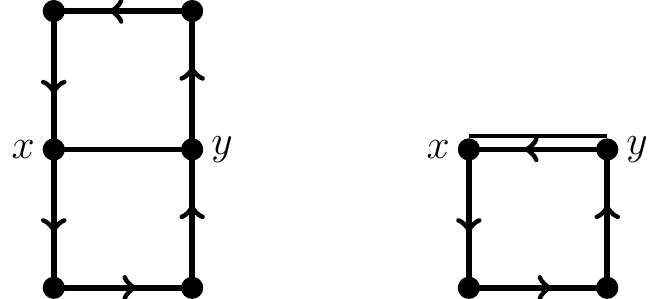}\caption{Two simple examples, where a closed baryon loop is connected to a single meson line at two sites $x$ and $y$. As discussed in the text, the weights of both diagrams are positive (in a nontrivial way).}
\label{fig_small_examples}
\end{figure}

Building up slightly more complicated configurations, consider a baryon loop connected to a line of unit mesonic $X$ at two sites $x$ and $y$. Fig. \ref{fig_small_examples} shows two simple examples of this kind. On all sites of the baryon loop except $x$ and $y$ the $\phi$-integrations discussed above force the flavor combinations $(1,2,3)$, $(1,2,4)$, $(1,3,4)$ and $(2,3,4)$ to traverse these parts of the loop separately. At $x$ and $y$, besides the positive $|d_{f_1,f_2,f_3}|^2$, the mixed terms already discussed come into play. The typical nonvanishing contribution reads
\begin{align}
 &\det(\phi^1|\phi^2|\phi^3)(x)
 \det(\phi^1|\phi^2|\phi^3)^*(y)\notag\\
 &\times 
 \det(\phi^1|\phi^2|\phi^4)(y)
 \det(\phi^1|\phi^2|\phi^4)^*(x)\notag\\
 &\times 
 \tr\big(\phi^4(x)\phi^4(y)^\dagger\phi^3(y)\phi^3(x)^\dagger\big)\notag\\
 &=\epsilon_{abc}\phi^1_a\phi^2_b\phi^3_c\phi^4_d(x)
 \,\epsilon_{ABC}\phi^{1*}_A\phi^{2*}_B\phi^{4*}_C\phi^{3*}_d(x)\notag\\
 &\times (x\rightarrow y)^*\,.
\end{align}
Now the $\phi^{1,2,3,4}(x)$-integrations are nonvanishing provided $A=a\,,B=b\,,c=d\,,d=C$, such that the field factor from site $x$ becomes positive, $|\phi_1|^2|\phi_2|^2|\phi_3|^2|\phi_4|^2$ (times Gaussian), with a positive prefactor $\epsilon_{abd}\epsilon_{abd}=6$. Such a factor appears from site $y$, too, and the total weight of these configurations are again positive.

In our opinion, these examples point at the positivity of all diagrammatic weights even for $N_f > 3$, when flavor dependent constraints, i.e., the conservation of each flavor current, are used at all sites. The full $\phi$-integration could then still be performed with Monte Carlo sampling. Using these finer constraints means to break down the flavor-summed exponents $\{j,\ldots,\bar{n}\}$ into flavor-dependent exponents through multinomials. The book-keeping of the nonvanishing terms becomes rather intricate, especially if higher occupations appear (which is determined by the dynamics of the system, its phases etc.). We leave this to future work. An alternative approach to project onto the relevant contributions are subsets \cite{Bloch:2011jx,Bloch:2013ara,Bloch:2015iha}. 

\section{Summary and outlook}

We have shown that, in the strong coupling limit, 
the sign problem in sQCD can, indeed, be solved 
for $N_f=1,2,3$ flavors. 
It is of further note that these lattice flavors 
correspond to the same number of flavors in the 
continuum theory. 
In the staggered fermion case this is a serious problem, 
as one staggered flavor generally corresponds to 
more than one flavor in the continuum. 
Furthermore the remaining doublers cannot be removed 
by the rooting trick in the diagrammatic formulation. 
Also using more than one staggered flavor seems to 
give rise to a serious sign problem even in the 
mesonic case in $U(3)$, cf. \cite{Fromm:2010lga}. 
In this respect, the scalar theory is much more feasible. 

For more than three flavors, $N_f>3$, further research 
is needed to decide whether the approach outlined at the end of 
section \ref{sec_sign_flavor} is viable and removes the sign problem. The discussed examples point to this conjecture. 

The case of $N_f>3$ is particularly interesting 
when one wants to go beyond strong coupling. 
Recent approaches which lend themselves easily 
to the formulation outlined in this paper are 
detailed in \cite{Budczies:2003za, Brandt:2016duy,Vairinhos:2014uxa}. 
The main goal in these references is to rewrite 
the gauge plaquette action in such a way as to 
make only single links appear in the new action, 
at the expense of introducing auxiliary bosonic field variables. 
The auxiliary fields are either scalar fields \cite{Brandt:2016duy} or 
matrix valued fields \cite{Vairinhos:2014uxa}. 
In particular, the additional scalar fields naturally increase the 
number of flavors to $N_f>3$. 
Thus, if the sign problem is, indeed, absent in that case, 
a diagrammatic simulation of full sQCD seems to be in reach. 

\vskip5mm
\noindent
{\bf Acknowledgments:} 
The authors are supported by the DFG (BR 2872/6-2 and BR 2872/7-1) and thank Jacques Bloch and Christof Gattringer for helpful discussions. FB is grateful to the Mainz Institute for Theoretical Physics (MITP) for its hospitality and its partial support during the completion of this work. 


%\bibliography{./scalar.bib}

\end{document}